# Title：Natural selection on human Y chromosomes

Authors：Chuan-Chao Wang[1], Li Jin[1,2,3], Hui Li[1,*]

**Affiliations:**

1. State Key Laboratory of Genetic Engineering and MOE Key Laboratory of Contemporary Anthropology, School of Life Sciences, Fudan University, Shanghai 200433, China
2. CAS-MPG Partner Institute for Computational Biology, SIBS, CAS, Shanghai, China
3. Institute of Health Sciences, China Medical City, Taizhou, Jiangsu, China

* Correspondence to: lihui.fudan@gmail.com

## Abstract

The paternally inherited Y chromosome has been widely used in population genetic studies to understand relationships among human populations. Our interpretation of Y chromosomal evidence about population history and genetics has rested on the assumption that all the Y chromosomal markers in the male-specific region (MSY) are selectively neutral. However, the very low diversity of Y chromosome has drawn a long debate about whether natural selection has affected this chromosome or not. In recent several years, the progress in Y chromosome sequencing has helped to address this dispute. Purifying selection has been detected in the X-degenerate genes of human Y chromosomes and positive selection might also have an influence in the evolution of testis-related genes in the ampliconic regions. Those new findings remind us to take the effect of natural selection into account when we use Y chromosome in population genetic studies.

## Keywords

Y chromosome; purifying selection; demographic history

## Introduction

In the field of anthropology, the uniparentally inherited Y chromosome has long been used to trace the paternal lineage of the populations and to understand differences in migration and population genetics between males and females, with additional advantages of small effective population size, low mutation rate, sufficient markers, and population-specific haplotype distribution (Jobling and Tyler-Smith, 1995; Jin and Su, 2000; Underhill et al., 2000). Many such population studies have rested on the assumption that all the Y chromosome markers in the non-recombination regions are selectively neutral (Jobling and Tyler-Smith, 2003). However, the Y chromosome is not free from variable phenotypic effects, e.g., the male infertility and Turner' syndrome (Ferlin et al., 2007; Hjerrild et al., 2008). Even if the particular Y chromosomal markers employed in a population study are non-functional, they may have been linked to detrimental or advantageous mutations or structural variations elsewhere on the chromosome. Because the bulk of Y chromosome does not undergo recombination, any positive or negative selection would affect the entire chromosome (Jobling and Tyler-Smith, 2003). Actually, among population geneticists, a debate has lasted for 30 years about whether the Y chromosome has been affected by natural selection. With the development of sequencing technology in recent years, Y chromosome sequence data has been gradually accumulated and enable us to address this complex issue step by step.

## Structure of Y chromosome

Human sex chromosomes, X and Y, evolved from a pair of homologous autosomes during the past 200–300 million years (Lahn and Page, 1999; Hughes et al., 2012). The human Y chromosome spans about 59 mega base pairs (Mbp) long, with 95% of its length inherited uniparentally from fathers to their sons without the recombination with X chromosome (Skaletsky et al., 2003). This majority region of Y chromosome has long been known as non-recombining region, or NRY, however, the discovery of abundant gene conversion in this region persuades us to rename it the male-specific region, or MSY (Rozen et al., 2003). The pseudoautosomal regions (PAR) flank on both sides of MSY, PAR1 and PAR2. PAR1 comprises 2.6 Mbp of the short-arm tips of both X and Y chromosomes in humans. PAR2 is located at the tips of the long arms, spanning 320 kbp. Known as relics of ancient homology between the X and Y chromosomes, PAR1 and PAR2 of Y chromosome can recombine with those on the X chromosome (Mangs and Morris, 2007).

The MSY is highly heterogeneous, consisting of two distinct types of sequence, heterochromatic and euchromatic. The heterochromatic portions mainly refer to a single 40 Mbp mass of heterochromatin on the long arm. The euchromatic portions comprise the X-transposed, X-degenerate, and ampliconic regions (Skaletsky et al., 2003). The X-transposed regions, which are 99% identical to DNA sequences in Xq21, originated from an X-to-Y transposition about 3–4 million years ago, after the divergence of human and chimpanzee. Only two genes have been identified within the 3.4 Mbp of X-transposed regions (Page et al., 1984; Skaletsky et al., 2003). The X-degenerate regions, which have a combined length of 8.6 Mbp, are suggested to be surviving relics of shared X–Y ancestry from ancient autosomes, sparsely populated with 16 single-copy gene or pseudogene homologs of X-linked genes. Most of X-degenerate genes are expressed widely throughout the body, however, the sex-determining *SRY* gene is found to be expressed predominantly in testis (Lahn and Page, 1997; Skaletsky et al., 2003). The ampliconic regions are composed largely of long duplicated sequences, spanning 10.2 Mbp. Most duplicated sequences are arranged in palindrome structures and form eight massive palindromes (P1-P8). Those regions exhibit the highest density of both coding and non-coding genes, with 60 genes in nine distinct multicopy gene families. In contrast to the widely expressed X-degenerated genes, the ampliconic genes and transcription units express primarily or exclusively in testis, thus, are of great significance for the evolution of male features (Rozen et al., 2003; Skaletsky et al., 2003; Hughes and Rozen, 2012).

## Low diversity of Y chromosome

A very low level of Y chromosome polymorphism has been reported since the middle of 1980s. During that time, Y-specific probes had been isolated from cosmid libraries and used in association with a set of restriction enzymes to search for male specific restriction fragment length polymorphisms (RFLPs) (Lucotte and Ngo, 1985; Ngo et al., 1986; Torroni et al., 1990; Oakey and Tyler-Smith, 1990; Malaspina et al., 1990). For instance, in the study of Malaspina et al. (Malaspina et al., 1990), 12 probes and 12 restriction enzymes were used to detect Y chromosome variations in 131 unrelated males. However, no polymorphic pattern was observed even in such a systematic search. A series of similar studies also conformed to the conclusion of very low diversity of human Y chromosome, compared with the autosomes and X chromosomes. Two possible reasons had been raised at that time to explain the low diversity (Malaspina et al., 1990). First, Y chromosomes are free from recombination, except for the very small pieces of PARs. However, recombination mechanism is in itself mutagenic. Thus, the paternally transmitted Y chromosomes are protected from mutations. Second, positive selection can reduce the level of polymorphism at the linked loci with favorable mutations, which is called genetic hitchhiking effect. In the case of Y chromosome, selectively driven fixation of an allele at any locus would result in the loss of polymorphism at the entire MSY (Malaspina et al., 1990).

However, most of the RFLPs mentioned above directed to the uncharacterized sequences in the nonrecombining region Yq11. In addition, the conclusion of low diversity of Y chromosomes was mainly drawn from the comparisons among Y chrosmomes, autosomes and X chromosomes, without using the interspecies divergence to correct for the mutation rate of the regions examined. Since the middle 1990s, direct sequencing of the well-characterized Y chromosomal regions of human and other primates has shed more light on the dispute about natural selection on human Y chromosome. In 1995, Dorit et al. sequenced a 729 bp intron of the *ZFY* region in 38 human samples and three nonhuman primates-chimpanzee, gorilla, and orangutan. Interspecies comparisons show that variable sites are distributed throughout this *ZFY* intron. However, no variation has been detected in the worldwide male human samples. Dorit et al. suggested a very recent common ancestor for human males (270,000 years with 95% confidence interval: 0 to 800,000 years) (Dorit et al., 1995). The large interval might imply the possible bias in the time estimation. Nevertheless, the lack of polymorphism at *ZFY* is probably a result of recent selective sweep. At the same time, Peter Goodfellow (Whitfield et al., 1995) only found three substitutions in humans by sequencing 18.3 Kbp of the sex-determining *SRY* region from five humans and one chimpanzee. Unlike the above other studies, Goodfellow compared the Y chromosome variations with the maternal inherited mitochondrial DNA (mtDNA). The mean human-human distance of Y chromosomes is 250 times smaller than that of the mtDNA. After correcting for the different mutation rates of Y chromosome and mtDNA, Goodfellow generated a very recent coalescence time for Y chromosmes (37,000~49,000 years), which is still 3 to 10 times smaller than that of mtDNA (120,000~474,000 years). The lower Y chromosome diversity compared with the mtDNA might support the suggested hitchhiking effect caused by selected sweep of an advantageous Y chromosome. However, the demographic history of human might also have influence on the genetic patterns. For instance, the cultural practice of polygyny could let a small number of males to have a disproportionately large number of offspring. Actually, people started to take the effective population size into account when searching for the possible natural selection on the human Y chromosomes. However, at almost the same time, Michael F. Hammer's study (Hammer, 1995) of Y Alu polymorphic (YAP) locus rejected the explanation that selection had reduced the diversity of human Y chromosome. Hammer sequenced 2.6 Kbp of YAP regions from 16 human and four chimpanzee Y chromosomes. He observed an average of one nucleotide difference between two randomly chosen YAP loci. The HKA test was used to search for selection at this locus. The within human variation ($\pi$) to DNA divergence between human and chimpanzee (*D*) at the YAP locus is almost the same as the $\pi:D$ ratios of mtDNA COII and ND4-5. Furthermore, the $\pi:D$ ratios of YAP and β-globin is not significantly different from the quadruple difference under a neutral model, which rejected the possible selective sweeps on the human Y chromosomes.

The above three studies have a common defect, the small sample size. There are also many ongoing debates about whether the comparisons among Y chromosome, mtDNA, and autosomes are the appropriate ways to evaluate the Y chromosome diversity. In 1999, Damian Labuda sequenced the last intron of Y chromosome *ZFY* gene and its X homolog *ZFX* to address those disputes. Only one variant was found within the 676-bp *ZFY* intron in 205 world-wide males. However, 10 variable sites were detected in 1089 bp of the ZFX region from 336 X chromosomes. The diversity of *ZFX* intron was higher than that of *ZFY*, but lower than that of neutrally evolved genomic region. Although the interspecies divergence in *ZFY* and *ZFX* was also reduced, the HKA and Tajima tests did not reject neutrality. It is possible that the very low diversity at the ZFY locus might reduce the power of HKA and Tajima tests. In addition, selection might be more difficult to detect in the recent expanding human populations (Jaruzelska et al., 1999). Then, the next question comes to whether population demographic history has influenced the Y chromosome diversity and the possible selection process, if so, to what extent?

## Influence of demographic history

Since the late 1990s, denaturing high-performance liquid chromatography (DHPLC) method has been used to detect the single nucleotide polymorphisms (SNPs) in the single-copy regions of MSY (Underhill et al., 1997, 2000). During the last ten years, a robust genealogical tree of human Y chromosomes based on numerous stable SNPs has been built, permitting inference of human population demographic history (Karafet et al., 2008; Yan et al., 2011).

Different demographic histories of male and female might influence the interpretation of the observed difference among paternal Y chromosome, maternal mtDNA, and autosomes, for instance, the sex-biased migration. Sex-biased migration refers to a higher female migration rate in human populations (Seielstad et al., 1998). A series of studies have revealed higher $F_{ST}$ values and lower diversities for the SNP and STR data of MSY than mtDNA and autosomes within or among world-wide populations, which indicates that Y chromosomes tend to be more localized geographically (Seielstad et al., 1998; Oota et al., 2001; Nasidze et al., 2004; Destro-Bisol et al., 2004; Wen et al., 2004; Wood et al., 2005). Most interesting is Stoneking's observation (Oota et al., 2001) that the $F_{ST}$ value of mtDNA in matrilocal population of Northern Thailand is more than two times higher than that of MSY, thus, this discrepancy is probably caused by residence patterns. The majority of human societies are patrilocal, in which a married couple tends to reside with or near the husband's parents. As a result, women move more frequently than do men between populations or groups, leading to the between-population differences for the paternal Y chromosome is greater than that of the maternal mtDNA (Stoneking, 1998). The patrilocality might be reasonable to explain the sex-biased genetic difference on a local scale. But now comes the question, how can patrilocality shape the continental or even global-scale sex-biased genetic patterns observed by Seielstad et al. (Seielstad et al., 1998)? Actually, there has been a debate about whether the SNPs in the genealogical tree of Y chromosome could be used for diversity inference. SNPs are usually ascertained in small numbers of males and then genotyped in much larger population samples. This nonrandom sampling of SNPs can result in an ascertainment bias, which might overestimate the diversity in the populations in which those SNPs are discovered (Jobling, 2012). Technological advances in sequencing now allow access to very large amounts of genetic data, which is eliminating this bias. For example, Michael F. Hammer directly compared 6.7 kb of Y-chromosome Alu region and 770 bp of the mitochondrial CO3 gene in 389 individuals from ten globally diverse human populations to see whether the global-scale patterns of sex-biased migrations really exist. Hammer found that within-continent and between-continent genetic differentiations for Y chromosome and mtDNA were all similar, which suggest broader-scale genetic patterns might not always be caused by residence patterns. Furthermore, the between-population variations in Hammer's data for Y chromosome and mtDNA were strongly and significantly correlated. This correlation, as Hammer suggested, might indicate that demographic history rather than natural selection is the primary determinant of genetic distance between the surveyed populations (Wilder et al., 2004a).

When we use demographic history rather than natural selection to explain the discrepant genetic pattern of Y chromosome and mtDNA, we have to be very cautious with the effective population sizes of male and female. Polygyny, higher rates of male mortality, or a greater variance in male lifetime reproductive success can decrease the effective population size of males, which subsequently reduce the Y chromosome diversity within populations relative to the mtDNA and autosomes (Seielstad et al., 1998; Wilder et al., 2004b).

Thus, a question posed to us is how to coordinate the relationship among sex-biased migration, natural selection, and effective population size. Hammer tried to address this issue by sequencing 26.5 Kbp of noncoding DNA from MSY and 782 bp of mitochondrial *Cox3* in a sample of 25 Khoisan, 24 Mongolians, and 24 Papua New Guineans (Wilder et al., 2004b). Tajima's D and Fu and Li's D* test were carried out to detect the possible deviation from neutrality. Only the Fu and Li's D* for the mtDNA in Mongolian is more negative than expected, and Fu and Li's D* for the MSY in Khoisan is slightly greater than expected, however, neither of the values were statistically significant. Furthermore, Hammer and his colleagues first assumed that the effective populations size for male and female were equal, and then performed coalescent simulations of bottlenecked genealogies under a variety of scenarios in Khoisan to see if a selective sweep had shaped difference in the MSY and mtDNA genealogies. However, those simulations, regardless of strong bottleneck (a 100-fold reduction in population size for 100 generations), weak bottleneck (10-fold size reduction for 100 generations) or bottlenecks with population contraction, all failed to produce results that are compatible with the observed levels of variation in Y chromosomes of Khoisan (Wilder et al., 2004b). Hammer concluded that a higher female effective population size rather than natural selection had caused the genetic difference between Y chromosome and mtDNA (Wilder et al., 2004b). Another factor that led Hammer to the foregoing conclusion is that mtDNA is likely to be subject to stronger purifying selection acting against deleterious mutations than MSY, if purifying selection really has an influence on mtDNA and Y chromosome. The reason is that mtDNA has a background mutation rate that is nearly 10 times higher than that of the MSY and mtDNA has much less redundancy as to gene content than the MSY (Wilder et al., 2004b). In addition, it is worthy to note that mtDNA mutations might be associated with male infertility. It has been reported that even a modest reduction in power output by mtDNA may reduce sperm mobility in human, but symptoms of disease are generally not observed until the function of mtDNA is reduced by 80% or more (Gemmell and Sin, 2002). That is to say mutations that might lead to the male infertility probably have very little effect on female fitness. In recent years, a lot of high-quality whole genome datasets have been available with the application of next-generation sequencing technology. In 2013, Sayres et al. used complete genomic sequence data from 16 African and European males to do genome-wide analyses in searching for the possible effect of natural selection. They have found the diversity of Y chromosome is extremely low, which is even an order of magnitude lower than the neutral expectation of one-quarter the diversity of autosomes. In contrast, the diversity of mtDNA is not reduced compared to expectations under neutrality. In their extensive simulations, they used the expected values under an equal male/female ratio for X/Autosome ratio (0.75) and for Y/Autosome and mtDNA/Autosome (0.25) as reference. Reduction in the male effective population sizes actually decreases the expected diversity of Y chromosome, however, the reduction required to explain the observed Y chromosome data, predicts levels of autosomes, X chromosomes and mtDNA diversity that are not consistent with the reference ratio of these genetic systems. Sayres et al. suggested that natural selection had reduced the diversity of human Y chromosomes. Furthermore, they performed forward simulations to confirm the role purifying selection had played in reducing the Y chromosome diversity. Their results also suggested selection might act on the ampliconic regions of Y chromosome (Sayres et al., 2013). As we have mentioned above, the ampliconic regions are responsible for the development of testis, their prediction makes some sense at this point.

Another complex question is that possible natural selection acting on Y chromosome has been difficult to confirm or exclude in expanding populations. We tend to consider the expansions of some specific Y chromosomal lineages as results of social selection rather than natural selection.

For instance, the expansion and world-wide distribution of Y chromosome haplogroup C3* star-cluster have been suggested to be associated with specific social prestige-the increased reproductive fitness of Genghis Khan (Zerjal et al., 2003). Similarly, haplogroup C3c-M48 was suggested to be spread by the Qing Dynasty Manchurian nobility (Xue et al., 2005). Another case, haplogroup O3a1c-002611 is one of the three main subclades of East Asia specific haplogroup O3-M122, accounting for about 16% of the Han Chinese. The late Neolithic expansion of O3a1c-002611 was suggested to be attributed to prosperity of agriculture (Wang et al., 2013). However, positive selection can also increase the frequency of a Y chromosomal lineage with advantageous mutations more rapidly than others. The question is how to distinguish the signal of possible selection from that of population expansion.

## Comparison with Y chromosomes of other primates

Population demographic history really poses a big problem for population geneticists to detect the possible selection on human Y chromosomes. However, the successful sequencing of complete Y chromosomes of human, chimpanzee, and rhesus macaque in David Page's lab has provided empirical data that enables us to directly search for the answer at evolutionary scale.

The Y chromosomes of human and rhesus have five evolutionary strata, each probably duo to an inversion to suppress the recombination between the Y chromosome and X chromosome. The oldest stratum dates back over 240 million years (Myr), and the youngest stratum originated only 30 Myr ago. Within the oldest four strata, the human and rhesus possess precisely the same 18 ancestral genes, meaning no gene were lost in those four strata during the past 25 Myr after the split of the human and rhesus. Furthermore, 17 of the ancestral 18 genes have a ratio of nonsynonymous substitution rate to synonymous substitution rate (dN/dS) that is less than one, of which 11 are statistically significant. This strict evolutionary conservation of ancestral genes on Y chromosomes is attributed to purifying selection (Hughes et al., 2012).

However, the MSY of chimpanzee has lost six genes in the 6 Myr since the split of the human and chimpanzee, but it contains twice as many massive palindromes in ampliconic regions as the human MSY (Rozen et al., 2003). As we have mentioned in the beginning, the genes in the ampliconic regions have testes restricted or exclusive expression patterns and critical functions in sperm production. The ampliconic regions of chimpanzee might be subject to powerful positive selection driven by fierce sperm competition (Hughes et al., 2010). In contrast, no X-degenerate gene loss has occurred during the last 6 Myr of human evolution. Furthermore, the human–chimpanzee interspecies divergence in coding regions of X-degenerate genes is significant less than that in introns, indicating purifying selection has played a vital role in maintaining the function of X-degenerate genes during recent human evolution (Hughes et al., 2005).

Then we can't help to ask if the purifying selection has continued to affect the human Y chromosomes in the even more recent period, such as out of Africa since 100 thousand years ago? David Page and his colleagues again focused on the X-degenerate genes to address this question. They sequenced the 16 X-degenerate single-copy genes and 5 single-copy pseudogenes in 105 male samples representing 47 world-wide lineages of the Y chromosomal genealogy. They found that nonsynonymous nucleotide diversity is significantly lower than diversity at synonymous sites, in introns, and in pseudogenes. However, the nucleotide diversity of the introns among the Y chromosomes studied was about an order of magnitude lower than that in the rest of the genome, suggesting other demographic factors probably have reduced the effective population size of males (Rozen et al., 2009).

## Conclusion

The long-going debate has come to an end with the solid evidence of natural selection on Y chromosomes. Purifying selection has maintained the strict evolutionary conservation of ancestral X-degenerate genes on Y chromosomes. Positive selection might play a role in the evolution of testis-related genes in the ampliconic regions. This new finding strongly suggests the natural selection on human Y chromosomes will be a vital parameter to consider when using Y chromosome data in population genetic studies.

## Acknowledgement

This work was supported by the National Excellent Youth Science Foundation of China (31222030), National Natural Science Foundation of China (31071098, 91131002), Shanghai Rising-Star Program (12QA1400300), Shanghai Commission of Education Research Innovation Key Project (11zz04), and Shanghai Professional Development Funding (2010001).